\documentclass[a4paper,aps,11pt,nofootinbib,showpacs]{revtex4}
\usepackage{amsmath}
\usepackage{amssymb}
\usepackage{amsfonts}
\usepackage[dvips]{graphicx}

\newcommand{\zs}{\zeta_{\rm s}}
\newcommand{\rs}{\rho_{\rm s}}

\begin{document}
\title{Stable Bound Orbits around  Black Rings
}
\hfill{OCU-PHYS 332}

\hfill{AP-GR 78}


\author{Takahisa Igata} 
\email{igata@sci.osaka-cu.ac.jp}
\author{Hideki Ishihara}
\email{ishihara@sci.osaka-cu.ac.jp}
\author{Yohsuke Takamori}
\email{takamori@sci.osaka-cu.ac.jp}
\affiliation{%
 Department of Mathematics and Physics,
 Graduate School of Science, Osaka City University,
 Osaka 558-8585, Japan}

\begin{abstract}
We examine bound orbits of particles around singly rotating black rings. 
We show that there exist stable bound orbits in toroidal spiral shape 
near the \lq axis\rq\ of the ring, 
and also exist stable circular orbits on the \lq axis\rq\ as special cases.  
The stable bound orbits can have arbitrary large size if the thickness of the ring 
is less than a critical value.
\end{abstract}

\maketitle

\section{Introduction}

Recently, motivated by modern unified theories, gravity in higher dimensions has 
attracted much interest. 
In particular, a lot of works are devoted to higher-dimensional 
black holes (see for a review \cite{LivingReview}). 
The rotating black hole solutions of arbitrary dimensions in vacuum were 
obtained by Myers and Perry \cite{Myers:1986un}, 
and the singly rotating black ring solutions in five dimensions were derived by 
Emparan and Reall \cite{Emparan:2001wn}. 
It is striking that the black ring solutions reveal that a black hole in vacuum is 
not specified only by its mass and angular momenta, i.e., 
higher-dimensional generalization of the black hole uniqueness does not hold in the 
four-dimensional form. 
After this discovery, rich varieties of black rings are found by many authors 
\cite{Black_Rings}. 

One of the most important step to study the black holes 
is the investigation of geodesics in the black hole geometry. 
It was shown that the geodesic equations are separable for Myers-Perry 
black holes in arbitrary dimensions \cite{Frolov:2003en}. 
In black ring geometries, geodesics are extensively studied 
in refs.\cite{Hoskisson:2007zk, Durkee:2008an}, 
and it was reported there that the black ring spacetimes hardly admit 
separability of general geodesics. 

It is an interesting question whether a gravitational object has bound orbits of particles. 
It would seem that the existence of stable circular orbits is a characteristic 
property of the four-dimensional gravity. 
For simplicity, let us consider the Schwarzschild metric in 
$n$ dimensions \cite{Tangherlini:1963bw}, 
\begin{align}
	ds^2 &= -\left(1-\frac{M}{r^{n-3}}\right)dt^2 
		+ \left(1-\frac{M}{r^{n-3}}\right)^{-1}dr^2 
		+ r^2 d\Omega_{n-2}^2, 
\end{align}
where $M$ is the parameter related to the mass of the black hole, 
and $d\Omega_{n-2}^2$ is the metric of a unit $(n-2)$-dimensional sphere. 
The effective potential for a free particle in this metric is 
\begin{align}
	V_{\rm eff}(r) = \frac{L^2}{2mr^2}  
	-\frac{mM}{2r^{n-3}} -\frac{ML^2}{2mr^{n-1}} + \frac12 m, 
\label{V_Sch}
\end{align}
where $m$ and $L$ are the mass and the total angular momentum of the particle, respectively. 
In the four-dimensional case, $n=4$, the first term, centrifugal potential, and 
the second term, gravitational potential, can make a local minimum which corresponds 
to a stable circular orbit, 
while there is no stable circular orbit in the case $n \geq 5$ because the effective 
potential has no local minimum\footnote{
In five-dimensional black holes with Kaluza-Klein type \cite{KKBH},
there exist stable circular orbits \cite{Matsuno:2009nz}. 
Even for asymptotically flat case, stable bound states of Nambu-Goto strings 
are possible \cite{TSS}. 
}. 
The absence of stable circular orbit was also shown 
in the five-dimensional Myers-Perry black holes \cite{Frolov:2003en}.

In the present article, we investigate stable bound orbits around the black ring 
metrics in five dimensions. 
In the vicinity of thin black rings, the boosted black strings 
mimic the geometry of the rings.   
Then, one expects that there exist stable bound orbits 
near the black ring horizon in spiral shape 
as in the black string geometry. 
On the other hand, the gravitational field of the black ring in far region 
is described by the five-dimensional Schwarzschild metric, approximately. 
Then, one would expect that there is no stable circular orbit bound 
by the black rings in this region. 
Contrary to this intuition, we show that 
there exist stable bound orbits of which size can be larger 
than the size of black ring. 
These are stable toroidal spiral orbits in general, and 
stable circular orbits as special cases. There exist stable bound orbits 
of arbitrary large size if the ring thickness is smaller than a critical value. 

\section{Geometry of Black Rings}

We consider the metric of black ring in the form
\begin{eqnarray}
	ds^2 &=& - \frac{F(y)}{F(x)}
	\left(dt-CR\frac{1+y}{F(y)} d\psi \right)^2
\cr
	&&+\frac{R^2}{(x-y)^2}F(x)
\left[
	- \frac{G(y)}{F(y)}d\psi^2 - \frac{dy^2}{G(y)}
	+ \frac{dx^2}{G(x)} + \frac{G(x)}{F(x)}d\phi^2
\right],
\label{BlackRing}
\end{eqnarray}
where
\begin{align}
	&F(\xi)=1+\lambda\xi, \quad G(\xi)=(1-\xi^2)(1 + \nu \xi),
\cr
	&C=\sqrt{\lambda(\lambda-\nu)\frac{1+\lambda}{1-\lambda}}, 
\end{align}
and the ranges of the coordinates $x$ and $y$ are 
\begin{align}
	-\infty \leq y \leq -1, \quad
	-1 \leq x \leq 1. 
\end{align} 
The parameter $R$ denotes the radius of the ring, and 
the parameters $\nu$ and $\lambda$ in the range
\begin{align}
	0<\nu\leq\lambda<1 
\end{align}
describe the thickness  of the black ring and rotation velocity 
in the $\psi$ direction, respectively. 
For the regularity at the two axes of rotation, $\lambda$ has to be chosen as 
\begin{align}
	\lambda=\frac{2\nu}{1+\nu^2}.
\end{align}
The event horizon in the topology S$^2 \times$ S$^1$ 
is given by $y=-1/\nu$.

\section{Particle Motion in Black Ring Spacetimes}

The Hamiltonian of a free particle with mass $m$ in a metric $g_{\alpha\beta}$ 
is generally given by
\begin{align}
	H &= \frac{N}{2}\left( g^{\alpha\beta}p_\alpha p_\beta + m^2\right), 
\end{align}
where $N$ is the Lagrange multiplier and $p_\alpha$ is the canonical momentum. 
Since $t$, $\phi$ and $\psi$ are cyclic coordinates in the case of 
black ring metric \eqref{BlackRing}, 
the momenta conjugate to these are constants of motion, say 
$p_t=-E, p_\phi= L_\phi$ and 
$p_\psi=L_\psi$. Then, we obtain the effective Hamiltonian 
\begin{align}
	H = \frac{N}{2}\left[
		g^{xx}p_x^2 +g^{yy}p_y^2
 		+ E^2 \left(U_{\rm eff}(x,y) + \frac{m^2}{E^2}\right)
		\right], 
\label{eq:Hamiltonian}
\end{align}
where 
\begin{align}
	U_{\rm eff}(x,y)
	 	=& g^{tt} + g^{\phi\phi}l_{\phi}^2
 		+ g^{\psi\psi}l_{\psi}^2 - 2g^{t\psi}l_{\psi}
\label{U_eff}
\end{align}
with
\begin{align}
	&g^{tt} = - \frac{F(x)}{F(y)} - \frac{C^2(x-y)^2(y+1)^2}{G(y)F(x)F(y)}, \quad
	g^{xx} = \frac{(x-y)^2}{R^2}\frac{G(x)}{F(x)},\quad
	g^{yy} = - \frac{(x-y)^2}{R^2} \frac{G(y)}{F(x)},
\cr
	&g^{\phi\phi} = \frac{(x-y)^2}{R^2 G(x)},\quad 
	g^{\psi\psi} = - \frac{F(y)(x-y)^2}{R^2 G(y) F(x)},\quad
	g^{t\psi} = - \frac{C(x-y)^2(y+1)}{R G(y) F(x)}, 
\end{align}
and $ l_\phi:=L_\phi/E,~ l_\psi:=L_\psi/E$. 

By variation with $N$, we get the Hamiltonian constraint condition
\begin{align}
	g^{xx}p_x^2 +g^{yy}p_y^2
 		+ E^2 \left(U_{{\rm eff}} + \frac{m^2}{E^2}\right)=0. 
\label{Hamiltonian_Constraint}
\end{align}
We regard the system of a free particle around the black ring as 
a particle moving in a two-dimensional curved space 
with the effective potential \eqref{U_eff}.  
The system is constrained by the condition \eqref{Hamiltonian_Constraint}.
For understanding the configuration of particle orbit, 
it is convenient to use  coordinates $(\zeta, \rho)$ defined by 
\begin{align}
	\zeta = R \frac{\sqrt{y^2-1}}{x-y}, \quad
	\rho = R \frac{\sqrt{1-x^2}}{x-y}, 
\label{New_Coordinates}
\end{align}
and the effective potential $U_{\rm eff}(\zeta,\rho)$ 
which is given by \eqref{U_eff} and \eqref{New_Coordinates}.

Stationary solutions of the Hamiltonian system \eqref{eq:Hamiltonian} 
are specified by
\begin{align}
	&U_{\rm eff}(\zeta,\rho) + \frac{m^2}{E^2}=0, 
\label{U0}
\\
	&\partial_\zeta U_{\rm eff}(\zeta,\rho)
	=\partial_\rho U_{\rm eff}(\zeta,\rho)= 0.
\label{DU0}
\end{align}
The position of the stationary solution in the $\zeta$-$\rho$ plane is determined by \eqref{DU0} and the 
particle energy $E$ for the stationary orbit is given by \eqref{U0}. 
The stability of the stationary solutions requires that the stationary 
points are local minimum of $U_{\rm eff}(\zeta,\rho)$. 

Typical shapes of the effective potential $U_{\rm eff}(\zeta,\rho)$ 
are shown in Fig.\ref{fig:Potential_Contour} 
as contour plots for the black ring metric with the thickness parameter $\nu=0.2$. 
There exist local minima of $U_{\rm eff}(\zeta,\rho)$ if $l_\phi$ and $l_\psi$ 
are chosen in a suitable finite range\footnote{
The range of $l_\phi$ and $l_\psi$ for stable bound orbits are shown 
in a separate paper \cite{IIT}.}. 
A local minimum point $(\zs, \rs)$, of which position depends on 
$l_\phi$ and $l_\psi$, implies a stable 
bound orbit of particle around the black ring. 
Each world line is tangent to a timelike Killing vector which is a linear 
combination of $\partial_t$, $\partial_\phi$ and $\partial_\psi$, and 
its projection on a timeslice makes 
a toroidal spiral curve on the two-dimensional torus, direct product of S$^1$ 
with radius $\zs$ and S$^1$ with radius $\rs$. 
Whether the toroidal spiral is closed depends on $l_\phi$ and $l_\psi$. 
A local minimum point $(\zs=0,\rs)$ 
on the $\rho$-axis, fixed two-dimensional plane 
of the rotation generated by $\partial_\psi$,
corresponds to a stable circular orbit with radius $\rs$. 
It is also possible that orbits with $l_\psi=0$ are toroidal spirals dragged 
by rotation of black ring, while orbits with $l_\psi\neq 0$ are circles.

\def\width_this_fig{70mm}
\begin{figure}[htbp]
\begin{tabular}{cc}
\begin{minipage}{0.5\hsize}
 \begin{center}
{(A) $l_{\phi} = 7.2$ and $l_{\psi} = 0.0$}\\
 \includegraphics[width=\width_this_fig]{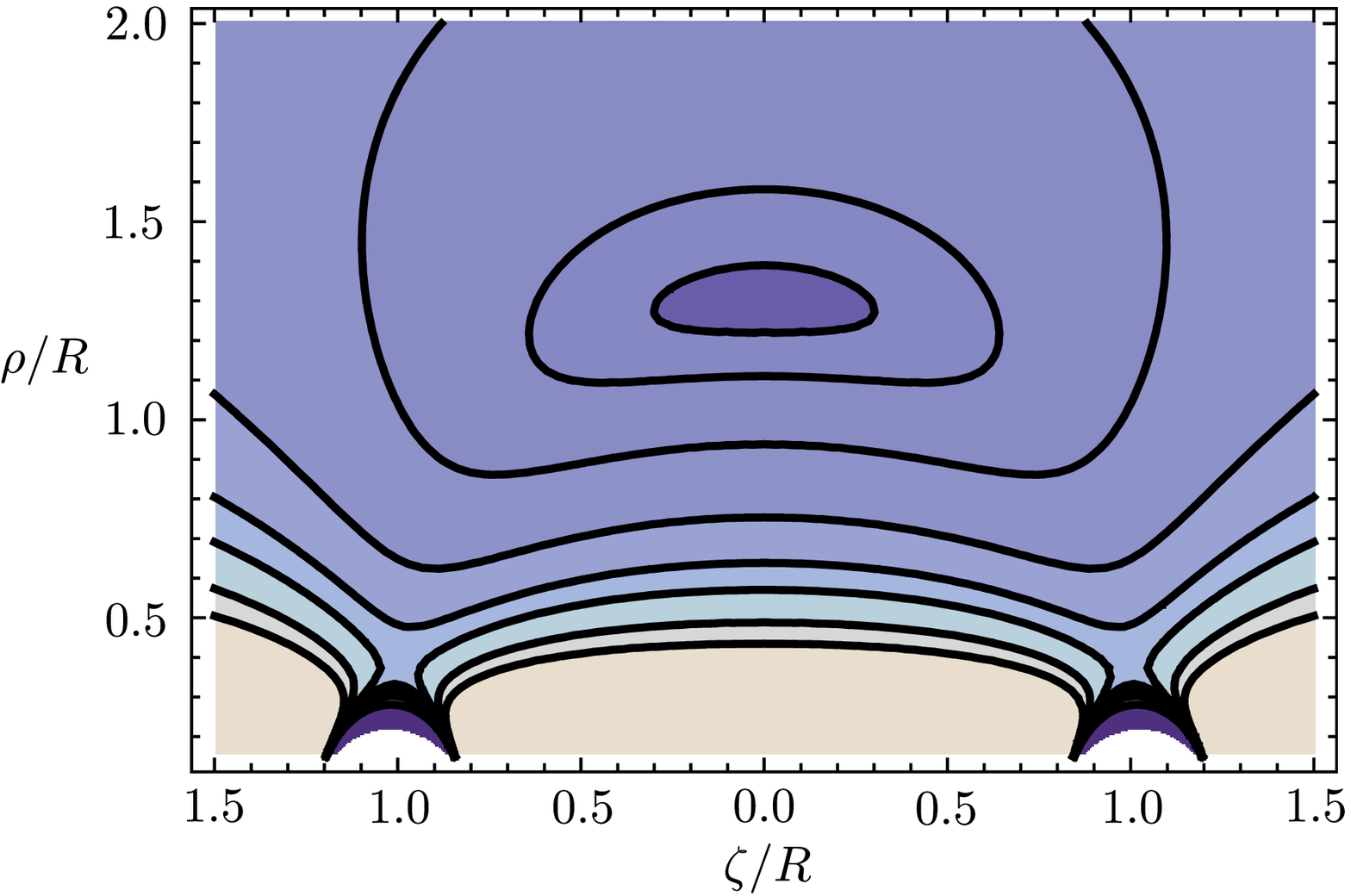}
\\
{(B) $l_{\phi} = 6.4$ and $l_{\psi} = 0.0$}\\
 \includegraphics[width=\width_this_fig]{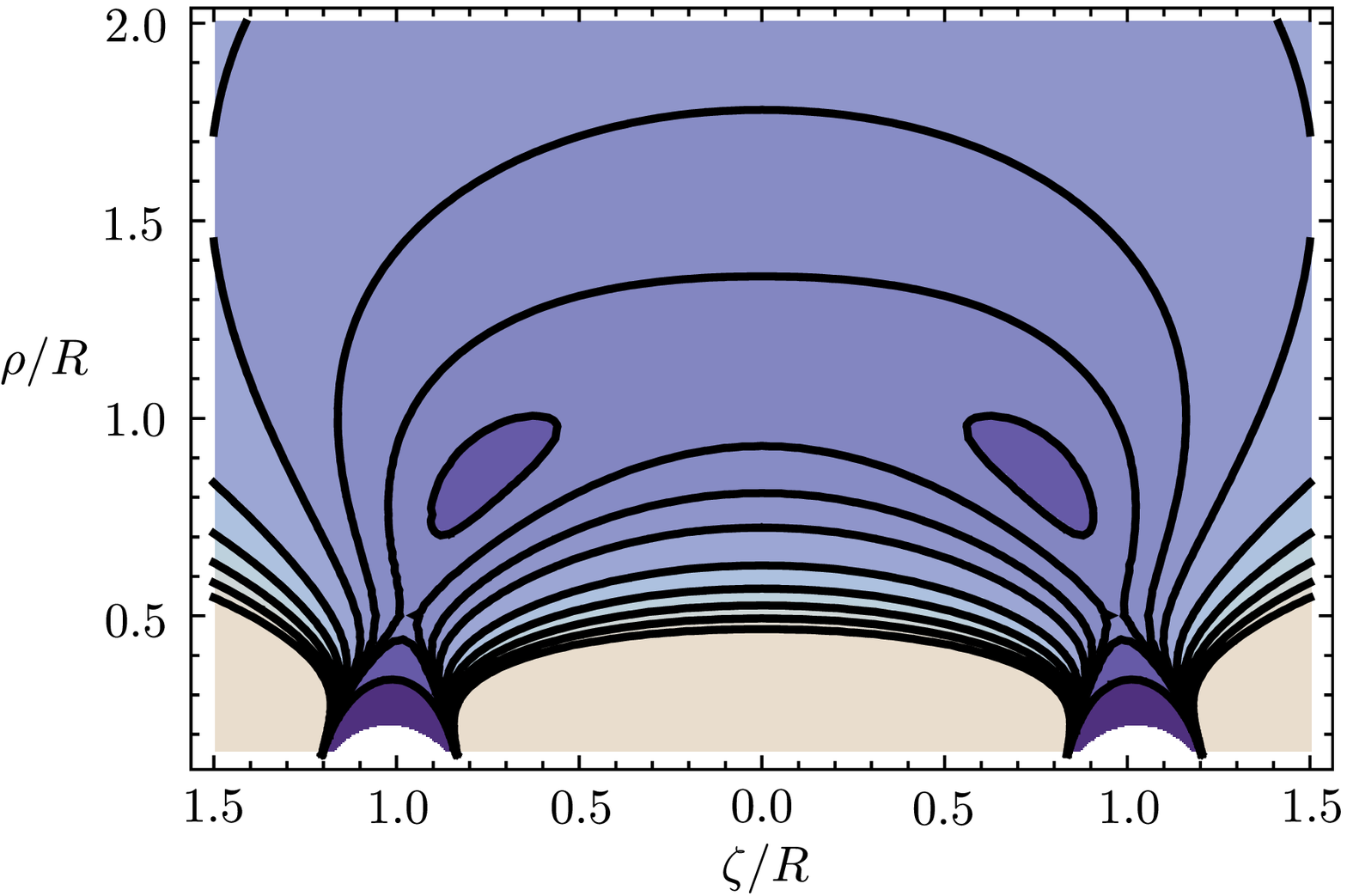}
\\
{(C) $l_{\phi} = 6.0$ and $l_{\psi} = 0.0$}\\
 \includegraphics[width=\width_this_fig]{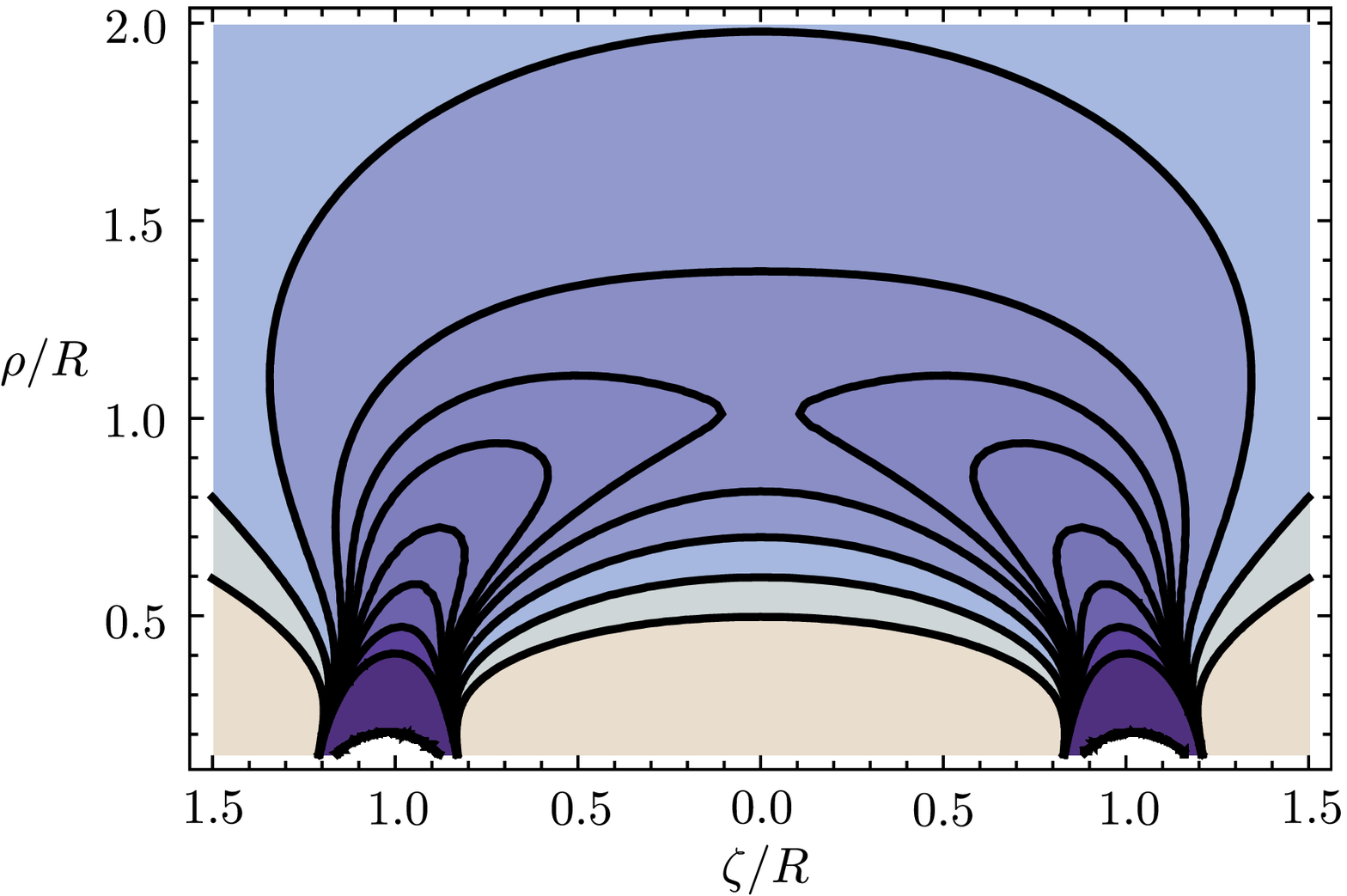}
 \end{center}
\end{minipage} 
\begin{minipage}{0.5\hsize}
 \begin{center}
 {(D) $l_{\phi} = 6.5$ and $l_{\psi} = 0.1$}\\
 \includegraphics[width=\width_this_fig]{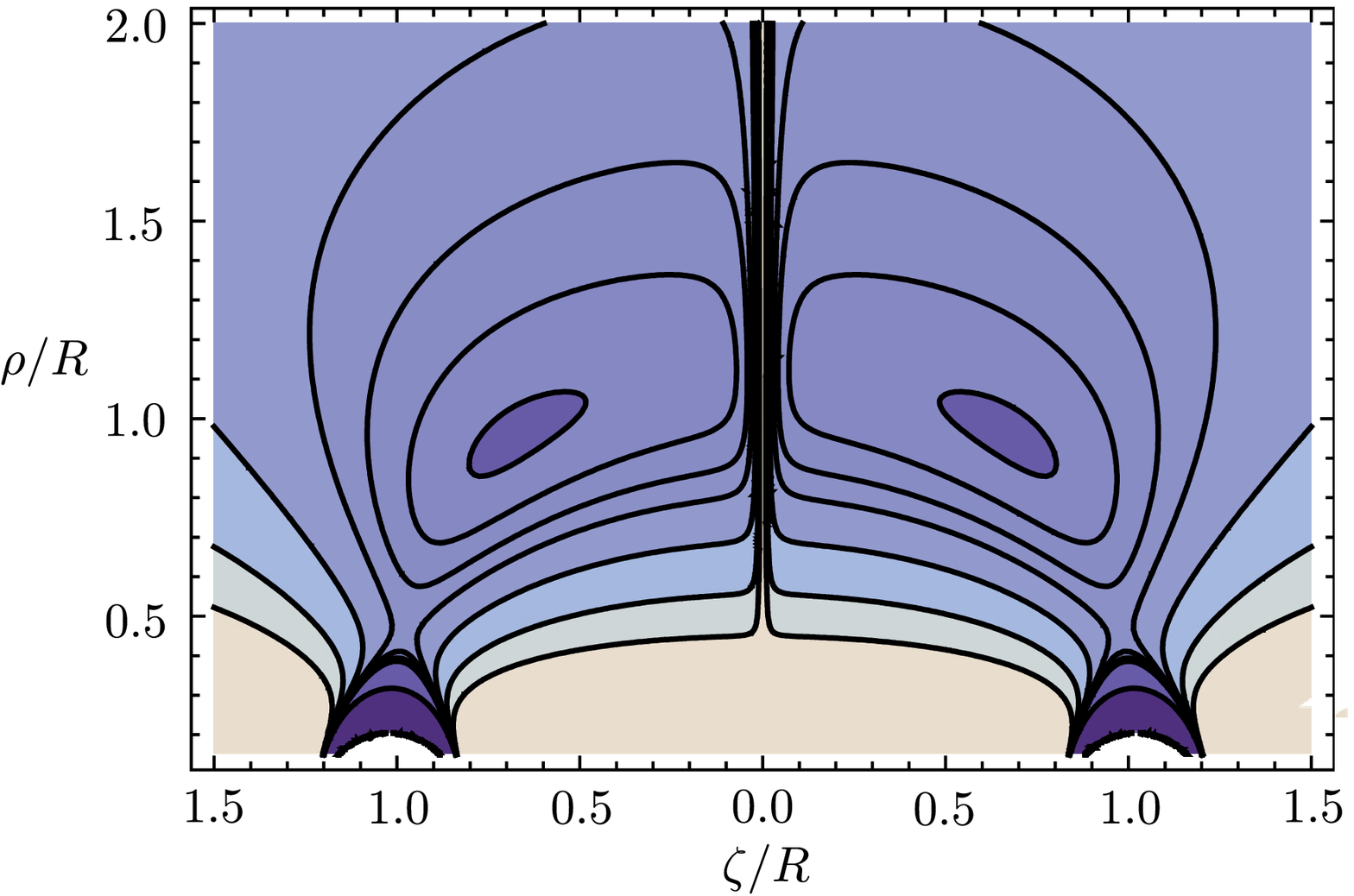}
\\
 {(E) $l_{\phi} = 6.5$ and $l_{\psi} = 1.5$}
\\
 \includegraphics[width=\width_this_fig]{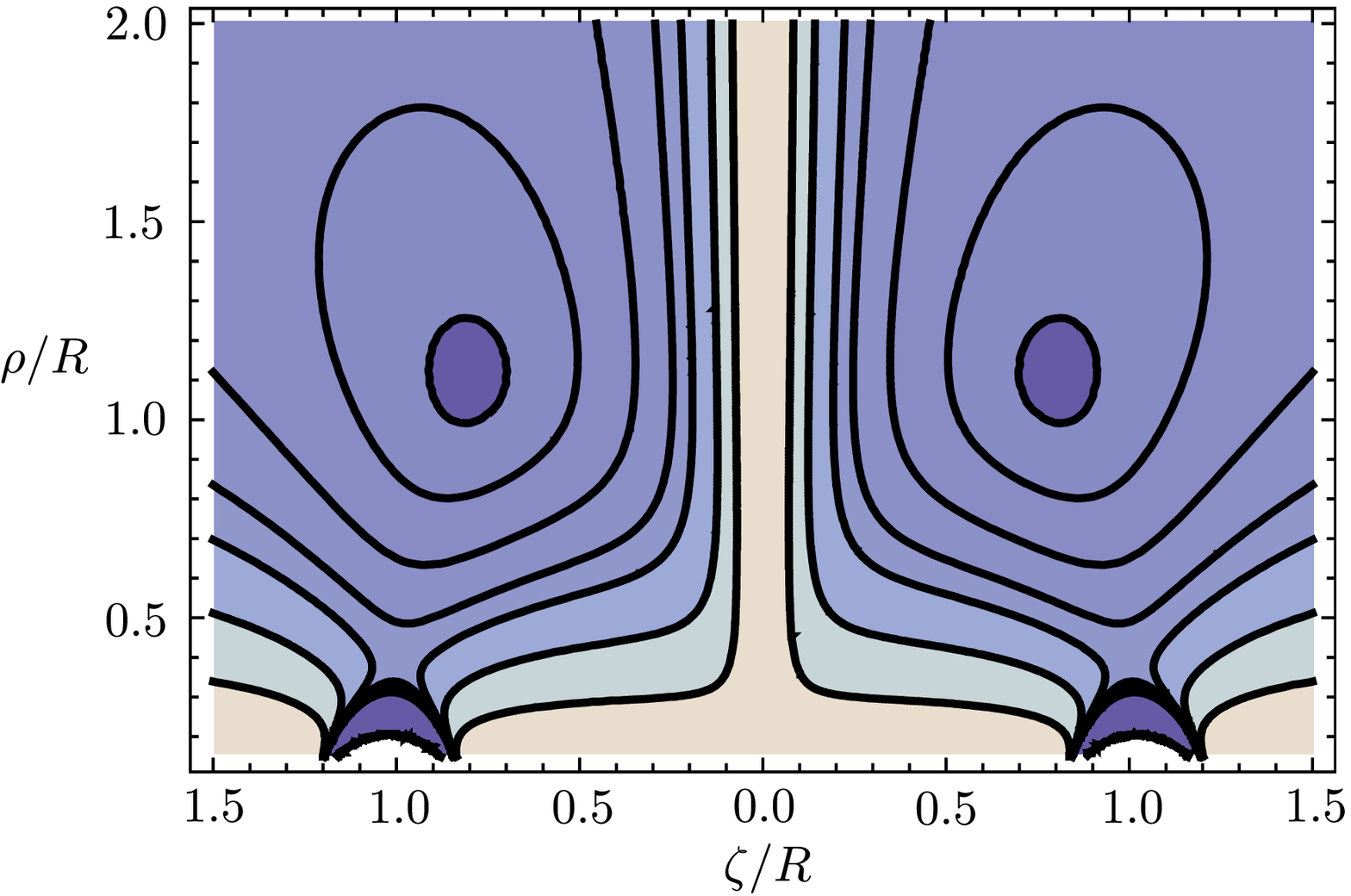}\\
 {(F) $l_{\phi} = 6.5$ and $l_{\psi} = 5.0$}
\\
 \includegraphics[width=\width_this_fig]{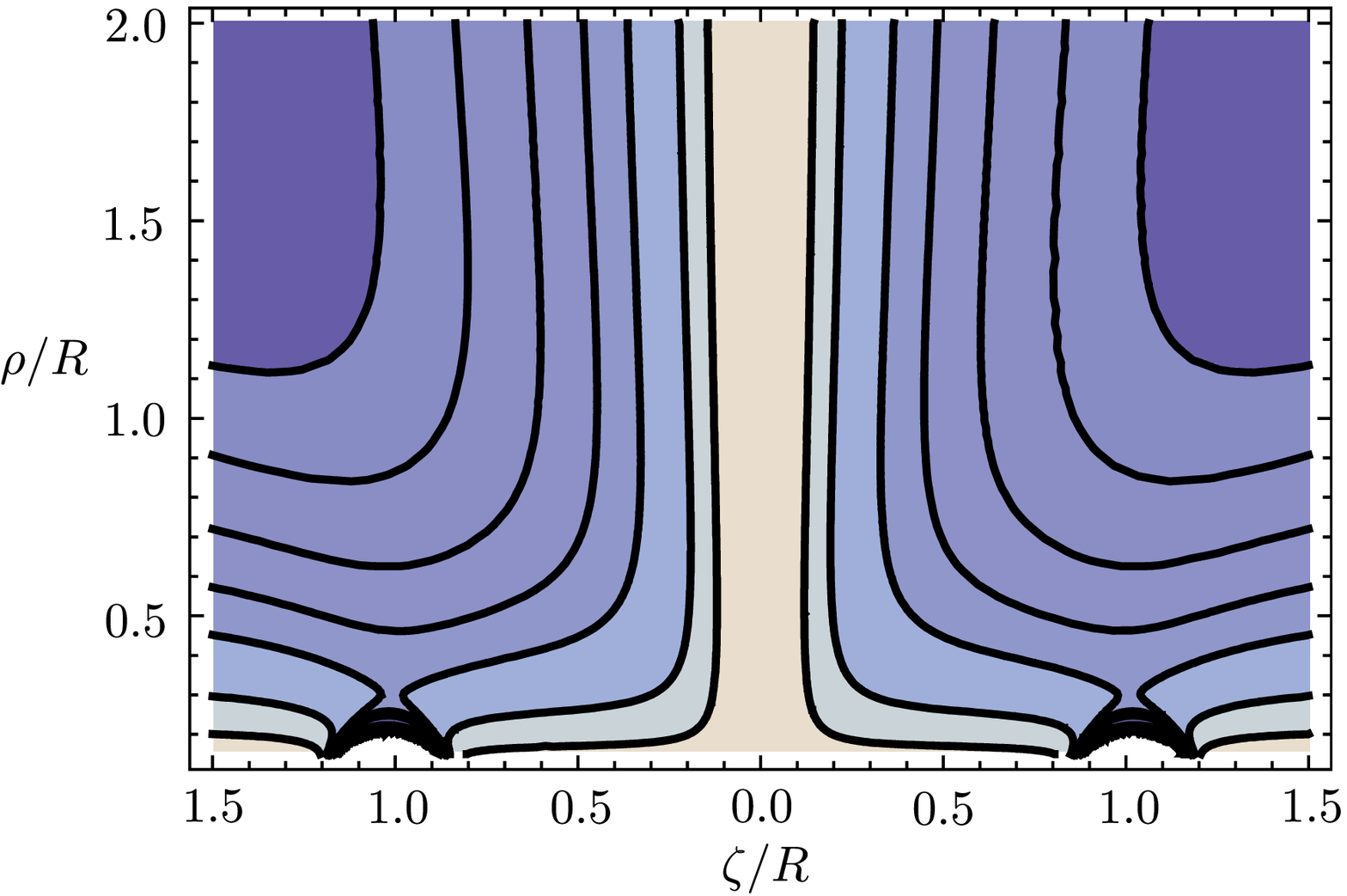}
 \end{center}
\end{minipage} 
\end{tabular}
 \caption{
Contour plots of $U_{\rm eff}(\zeta, \rho)$ in the case of the black ring with 
the thickness parameter $\nu = 0.2$. The horizontal axis is $\zeta/R$ and 
the vertical axis is $\rho/R$. The figures (A),(B),(C) are for 
$l_\psi=0$ cases. In (A) a local minimum exists on the $\rho$ axis, 
and in (B) two minima appear off the axis, while in (C) no minimum exists. 
The figures (D),(E),(F) are  for the cases of non-vanishing $l_\psi$. 
Potential barrier near the $\rho$ axis appears. 
Two local minima appear in (D) and (E), while no minimum in (F).
\label{fig:Potential_Contour}
\vspace{5mm}}
\end{figure}

\begin{figure}[tbp]
  \includegraphics[width=140mm]{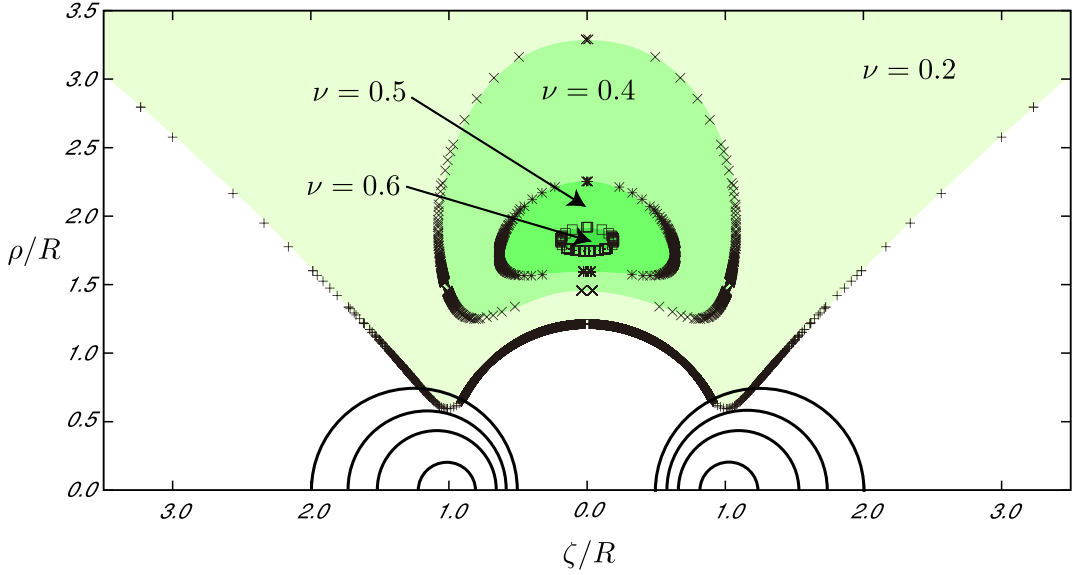}
 \caption{
Domains of stable bound orbits are superposed in the $\zeta$-$\rho$ plane 
as dark regions.The domain, in which potential minima exist, 
are obtained numerically 
for the black rings with $\nu=0.2, 0.4, 0.5, 0.6$.  
Half circles on the horizontal axis denote event horizons of the black rings 
with $\nu=0.2, 0.4, 0.5, 0.6$, from small semi-circle to large one. 
\label{fig:domain}
\vspace{5mm}
}
\end{figure}

For a fixed thickness parameter $\nu$, the local minima appear in a limited region in the $\zeta$-$\rho$ plane. 
By searching local minimum points numerically, 
we show that the domain of the stable bound orbits for the cases 
$\nu= 0.2, 0.4, 0.5,$ and $0.6$ in Fig.\ref{fig:domain}. 
There exist stable bound orbits on and near the axis of black ring, 
and not exists near the equatorial plane. 
There are two critical values of the thickness parameter $\nu$, say $\nu_\infty$ 
and $\nu_0$. 
The black rings with the thickness parameter in the range 
$0< \nu < \nu_\infty$ allow 
stable bound orbits in semi-infinite domains (see Fig.2 for the shape of domain).  
The black rings with $\nu_\infty < \nu \leq \nu_0$ 
admit the stable bound orbits in finite domains, 
and the rings with $\nu_0 < \nu <1$ have no stable bound orbit.

The critical values $\nu_\infty$ and $\nu_0$ can be obtained 
if we concentrate on stable circular orbits 
on the $\rho$-axis with $l_\psi=0$. 
By the regularity at the $\rho$-axis, it holds trivially that
\begin{align}
	\partial_\zeta U_{\rm eff}(\zeta=0, \rho) =0, 
\end{align}
then the stable circular orbits should satisfy the stationary condition
\begin{align}
	\partial_\rho U_{\rm eff}(\zs=0, \rs) =0 ,
\label{DU00}
\end{align}
and the stability conditions
\begin{align}
	\partial_\rho^2 U_{\rm eff}(\zs=0,\rs) >0 , \quad
	\partial_\zeta^2 U_{\rm eff}(\zs=0,\rs) >0 .
\label{DDU00}
\end{align}

First, we see the asymptotic form of the effective potential at large distance as
\begin{align}
	U_{{\rm eff}}(\zeta=0,\rho)  
	\simeq 
		-1-\frac{4 R^2 \nu -(1-\nu )l_\phi^2 }{(1-\nu)^2 ~~\rho^2} 
		+\frac{2 \nu R^2 \left(2R^2-l_\phi^2\right)}{(1-\nu)^2~~\rho^4},\label{U_asympt}
\end{align}
and
\begin{align}
	\partial_\zeta^2 U_{\rm eff}(\zeta=0,\rho)  
	\simeq 	\frac{4 R^2 \nu }{(1-\nu)^2 ~~\rho^4} 
	> 0.
\end{align}
Then, if two inequalities 
\begin{align}
		4 R^2 \nu -(1-\nu )l_\phi^2 >0 
\quad \mbox{and}\quad
		2R^2-l_\phi^2 >0
\end{align}
hold, all of stationary and stability conditions \eqref{DU00}-\eqref{DDU00} 
are satisfied at 
\begin{align}
	\rs^2 = \frac{4\nu R^2(2R^2-l_\phi^2)}{4R^2\nu-(1-\nu)l_\phi^2}. 
\end{align}
If $l_\phi^2$ approaches to $4 R^2 \nu/(1-\nu )$ 
in the range $2R^2>l_\phi^2$, the radius of the stable circular orbit 
$\rs$ becomes infinite. 
It can happen for the cases $0<\nu<\nu_\infty:=1/3$. 
The positive sign of the $\rho^{-4}$ term in \eqref{U_asympt} makes contrast to 
the negative sign of the $r^{-4}$ term for $n=5$ case in \eqref{V_Sch}. 
We note that the angular momentum $L_\phi=l_\phi E$ for the stable circular orbit 
is finite even if its radius is infinite. 

Secondly, we see that 
the other parameter $\nu_0$ gives the maximum thickness of the black ring which has 
the stable bound orbit. 
By inspecting $U_{\rm eff}$, 
we find that equations 
\begin{align}
	&\partial_\rho^2 U_{\rm eff}(\zs=0,\rs) =0 ,
\quad
	\partial_\zeta^2 U_{\rm eff}(\zs=0,\rs) =0 ,
\label{critical}
\end{align}
hold simultaneously for a value of thickness parameter $\nu=\nu_0$.
By solving the coupled algebraic equations \eqref{DU00} and \eqref{critical} 
for $\nu=\nu_0, l_\phi=l_0, \rs=\rho_0$, we have\footnote{
The critical value of $\nu_0$ is obtained in \cite{Hoskisson:2007zk}, but 
the meaning is different.}
\begin{align}
	\nu_0 
	=& \frac{13}{2}+\frac{1}{2} \left(145-24 \left(\frac{2}{3+\sqrt{41}}\right)^{1/3}
		+6\left(4(3+\sqrt{41})\right)^{1/3} \right)^{1/2}
\cr
	&\quad  -\left[\frac{145}{2}+6
   \left(\frac{2}{3+\sqrt{41}}\right)^{1/3} 
	-3 \left(\frac{3+\sqrt{41}}{2}\right)^{1/3} \right. 
\cr
	&\quad \left.	+\frac{1783}{2} 
	\left(145-24 \left(\frac{2}{3+\sqrt{41}}\right)^{1/3}
	+6\left(4(3+\sqrt{41})\right)^{1/3} \right)^{-1/2}\right]^{1/2}
\cr
	=& 0.65379\cdots , 
\\
	\frac{l_0^2}{R^2}	=& 
	\frac{3 \nu_0^2-8 \nu_0 +1}{\nu_0 -1}
	+\sqrt{\frac{ 6\nu_0 (\nu_0 ^2-8 \nu_0 +3 )}{\nu_0 -1}} ,
\\
	\frac{\rho_0^2}{R^2} 
	=&  \frac{ (1+\nu_0)^3 -2 (1-\nu_0)^2~ l_0^2/R^2 }
		{(1-\nu_0)^2~ l_0^2/R^2 -2 \left(1-\nu_0^2\right)}.
\end{align}
For the black rings with $0 < \nu < \nu_0$ we can find stationary points which 
satisfy the stability conditions \eqref{DDU00}, 
while the rings with $\nu_0 < \nu <1$ 
the conditions \eqref{DDU00} break at all stationary points.

In the parameter range $0<\nu<\nu_\infty$, the potential value of 
stable circular orbit satisfies
\begin{align}
	 -\frac{m^2}{E^2}=U_{{\rm eff}}(\zs=0,\rs)<-1
\end{align}
then, the binding energy $E_b :=m-E$ is positive. 
There appears one more critical value $\nu_+$ such that if $\nu_+ < \nu < \nu_0$ 
the binding energy $E_b$ of stable circular orbit is negative\footnote{
We check these properties for stable toroidal spiral orbits numerically 
up to numerical accuracy.}. 
This possibility is pointed out in ref.\cite{Hoskisson:2007zk}. 
In the case $\nu_\infty<\nu<\nu_+$, 
the sign of the binding energy depend on the parameter $l_\phi$.  
The critical parameter $\nu_+$ is obtained algebraically as
\begin{align}
	\nu_{+}
		&= \frac{-8(43-3\sqrt{177})^{2/3}+(86-6\sqrt{177})^{1/3}(-25+\sqrt{177})
		+2^{2/3} (-68+4\sqrt{177})}{-16(43-3\sqrt{177})^{2/3}
		+(86-6\sqrt{177})^{1/3}(-41+\sqrt{177})+2^{2/3}(-154+10\sqrt{177})}
\cr
		&= 0.52444\cdots .
\end{align}
In the limiting case of black ring $\nu=\nu_0$, we have the limiting 
circular orbit with the radius $\rho_0$ with the angular momentum parameter $l_0$. 
The potential value for the orbit becomes 
$-m^2/E^2=U_{\rm eff}(\zs=0,\rho_0) = 0$, 
that is, energy $E$ and angular momentum $L_\phi = l_0 E$ of the particle 
diverge. 
Namely, particles with almost light velocity can be trapped stably in a 
finite radius by the black ring with the thickness parameter slightly less 
than $\nu_0$.

\section{Discussions}

Black rings in vacuum cannot be distinguished from rotating black holes by 
their mass and angular momenta in general. 
However, existence of stable bound orbits, which is not possessed by 
five-dimensional black holes, is a unique property for black rings 
with the thickness parameter $\nu<\nu_0\approx 0.65379$. 
In addition to stable toroidal spiral orbits near black ring horizon, 
stable circular orbits whose radii are much larger than the ring radius can exist 
for thin ring with $\nu<\nu_\infty =1/3$. 
The leading order of gravitational potential at large distance 
is proportional to inverse square of distance in five dimensions, 
which is the same order of the centrifugal potential. 
The next order term, inverse fourth power of distance, is essential 
for the existence of stable bound orbits. 
The sign and amount of this term depend on the shape of gravitational source. 
Thin black rings cause the appearance of stable bound orbits 
at large distance.  
Near the stable bound orbits, there exist dynamical orbits bounded in finite regions. 

For the black rings with thickness parameter $\nu<\nu_\infty$, 
the binding energy of the bound orbit is positive. 
Then, a particle falling toward a black ring can be trapped in a stable bound orbit 
by energy loss as usual. 
In contrast, for the black rings with $\nu_+\approx 0.52444<\nu<\nu_0$, 
all bound orbits have negative binding energies. 
Though it would be difficult to raise a particle into the bound orbit with 
negative binding energy, 
during a violent process of black ring formation highly energetic particles 
would be trapped in stable bound orbits like a storage ring.

The absence of stable bound orbit of test particle around five-dimensional 
black holes would suggest absence of black hole binary system in five dimensions. 
In contrast, does the existence of stable bound orbits of particles around black rings suggest the existence of black hole-black ring bound system?

This work is supported by the Grant-in-Aid for Scientific Research No.19540305.

\end{document}